\documentclass[twocolumn,preprintnumbers,elsart]{revtex4}
\usepackage{eurosym}
\usepackage{makeidx}
\usepackage{amssymb}
\usepackage{amsmath}
\usepackage{mathrsfs}
\usepackage{graphicx}
\usepackage{dcolumn}
\usepackage{bm}
\usepackage[center]{subfigure}
\usepackage{color}
\usepackage{indentfirst}
\usepackage{array}
\usepackage{booktabs}
\usepackage{multirow}
\usepackage{upgreek}

\begin{document}

\title{Strongly anisotropic vortices in dipolar quantum droplets}
\author{Guilong Li$^{1}$}
\author{Zibin Zhao$^{1}$}
\author{Xunda Jiang$^{1}$}
\author{Zhaopin Chen$^{2}$}
\author{Bin Liu$^{1}$}
\email{binliu@fosu.edu.cn}
\author{Boris A. Malomed$^{3,4}$}
\author{Yongyao Li$^{1,5}$}
\email{yongyaoli@gmail.com}
\affiliation{$^{1}$School of Physics and Optoelectronic Engineering, Foshan University,
Foshan 528225, China\\
$^{2}$Physics Department and Solid-State Institute, Technion, Haifa 32000,
Israel\\
$^{3}$Department of Physical Electronics, School of Electrical Engineering,
Faculty of Engineering, Tel Aviv University, Tel Aviv 69978, Israel \\
$^{4}$Instituto de Alta Investigaci\'{o}n, Universidad de Tarapac\'{a},
Casilla 7D, Arica, Chile\\
$^{5}$Guangdong-Hong Kong-Macao Joint Laboratory for Intelligent Micro-Nano
Optoelectronic Technology, Foshan University, Foshan 528225, China}

\begin{abstract}
We construct strongly anisotropic quantum droplets with embedded vorticity
in the 3D space, with mutually perpendicular vortex axis and polarization of
atomic magnetic moments. Stability of these anisotropic vortex quantum
droplets (AVQDs) is verified by means of systematic simulations. Their
stability area is identified in the parametric plane of the total atom
number and scattering length of the contact interactions. We also construct
vortex-antivortex-vortex bound states and find their stability region in the
parameter space. The application of a torque perpendicular to the vorticity
axis gives rise to robust intrinsic oscillations or rotation of the AVQDs.
The effect of three-body losses on the AVQD stability is considered too. The
results show that the AVQDs can retain the topological structure (vorticity)
for a sufficiently long time if the scattering length exceeds a critical
value.
\end{abstract}

\maketitle

Quantum droplets (QDs), representing a novel form of quantum matter, have
drawn much interest in recent years \cite%
{Ferrier-Barbut2018,Guo2021,Luo2021,BAM2021,Bottcher2021,Chomaz2023,Ferrier-Barbut2016a,GEA2018,Skov2021,Yin2021,Yin2022,Wang2020,Yo2023,yaw2024,dlw2021,zh2024,henn2024}%
. These are droplets of an ultra-dilute superfluid maintained by the balance
between the mean-field (MF) and beyond-MF effects \cite%
{Petrov2015,Petrov2016}, the latter one being the Lee-Huang-Yang (LHY)
correction \cite{LHY,Jrgensen2018} to the MF nonlinearity induced by quantum
fluctuations. QDs have been experimentally observed in dipolar Bose-Einstein
condensates (BECs) \cite{Schmitt2016,Chomaz2016}, as well as in binary BECs
of nonmagnetic atoms, with quasi-2D \cite{Cabrera2018,Leticia2} and 3D \cite%
{Semeghini2018,collision,Salasnich}. On the contrary to that, experimental
realization of self-trapped BECs in free space solely through MF effect is
impossible due to the critical or supercritical collapse instability in the
2D and 3D settings, respectively \cite%
{Fibich1999,Berge1998,Kuznetsov2011,Pethick2002} (nevertheless, weakly
unstable quasi-2D \textit{Townes solitons} were experimentally created in a
binary BEC \cite{Townes1,Townes2}). 3D QDs in nonmagnetic condensates appear
in the isotropic form, whereas their shapes are anisotropic in dipolar BECs
\cite{Baillie2016,Wenzel2018} (a possibility to create stable isotropic QDs
in a 2D dipolar BEC was considered too \cite{NJP}). Note that stable
anisotropic quasi-2D fundamental (zero-vorticity) MF solitons (in the
absence of the LHY correction) can be created in the magnetic BEC with the
in-plane dipole polarization \cite{Vardi}, but they do not exist in the 3D form \cite{Lahaye2008}.

QDs are the subject of a vast research area, including Monte-Carlo
simulations \cite{Astra,Parisi2019,VC2020}, collective excitations \cite%
{Tyluki2020,Huhui2020,Baillie2017}, supersolids, \cite%
{Yongchang,Yongchang2021,Bttcher2019,Hertkorn2021,Sanchez-Baena2023,Scheiermann2023}%
, Borromean droplets \cite{Xiaoling2021}, etc. A particularly interesting
direction of the studies is embedding vorticity into the self-bound QDs. It
is well known that the creation of self-trapped vortices in the
multi-dimensional space is a challenging issue. The azimuthal instability,
which is induced by the underlying self-attraction, tends to split the 2D
vortex ring or 3D torus (\textquotedblleft donut") into fragments \cite{Malomed2022c,Michinel,Pego,Dumitru,Malomed2019}. This instability develops
faster than the collapse driven by the self-attraction. In QDs, the
splitting instability may be arrested by the competition between the MF
attraction and LHY self-repulsion, similar to the stabilizing effect of the
cubic-quintic nonlinearity in optics \cite{Michinel,Pego,Dumitru,Cid}.
Stable vortex QDs with the winding numbers (topological charge)\ up to $5$
and $2$ (at least) were predicted in 2D \cite{Li2018a} and 3D geometries
\cite{Kartashov2018}, respectively. Stable semi-discrete vortex QDs were
predicted in arrays of one-dimensional traps \cite{Zhang2019}. These results
indicate that the equilibrium state of the LHY-stabilized superfluid
provides a versatile platform for the creation of the stable self-bound
vortices \cite{cgh2024}.

\begin{figure}[h]
{\includegraphics[width=0.8\columnwidth]{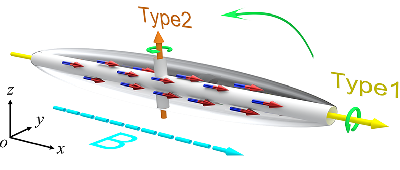}}
\caption{Possible relations between the vorticity axis and polarization of
atomic magnetic dipoles, which is fixed by magnetic field \textbf{B} along
the $x$ axis. Type 1: the vorticity parallel to the polarization (the same
as in Ref. \protect\cite{Cidrim2018}). This configuration is always
unstable. Type 2: the new configuration, which may be stable, with the
vorticity oriented perpendicular to the polarization.}
\label{example}
\end{figure}

The above-mentioned findings were produced for binary BECs of nonmagnetic
atoms. For the dipolar QDs, isotropic vortex modes have been reported, with
the vortex axis parallel to the polarization of atomic magnetic moments,
represented by \textquotedblleft Type 1" in Fig. \ref{example}. This
configuration is rotationally symmetric with respect to the vorticity axis,
but it is known to be unstable \cite{Cidrim2018}. The creation of \emph{%
anisotropic} vortex QDs in dipolar BECs and their stability is a challenging
problem. This problem is also relevant in studies of other nonlinear
systems, as no example of anisotropic vortex solitons in free space was
reported in any context. Very recently, stable vortex QDs were predicted in
a 2D dipolar setup \cite{Li2023}. However, this problem was not previously
addressed in the full 3D geometry.

In this Letter, we predict the existence of stable 3D strongly anisotropic vortex quantum droplets (3D-AVQDs) in the dipolar BEC with the magnetic dipoles polarized perpendicular to the vortex' axis, corresponding to the \textquotedblleft Type 2" configuration in Fig. \ref{example}. Note that this configuration is not a straightforward extension of its 2D counterpart. In particular, the introduction of the third direction may readily give rise to flexural instability along the axis perpendicular to the plane of the original 2D mode, thus destroying the stability of the original 2D vortices. Moreover, the instability can also induce abundant new dynamics, such as the self-rotation along the dipole orientation, which does not exist in their 2D counterparts. Therefore, the stability of the 3D AVQDs is a challenging issue.

The respective 3D LHY-amended Gross-Pitaevskii equation (GPE) is written as

\begin{gather}
i\hbar \frac{\partial }{\partial t}\psi =-\frac{\hbar ^{2}}{2m}\nabla
^{2}\psi +g|\psi |^{2}\psi +  \notag \\
\kappa \psi \int U_{\mathrm{dd}}(\mathbf{r}-\mathbf{r^{\prime }})|\psi (%
\mathbf{r^{\prime }})|^{2}d\mathbf{r^{\prime }}+\gamma |\psi |^{3}\psi +i%
\frac{\hbar }{2}\Lambda _{3}|\psi |^{4}\psi ,  \label{GP-LHY}
\end{gather}%
where $\hbar $ and $m$ are the Planck's constant and atomic mass, $g=4\pi
\hbar ^{2}a_{s}/m$ with $a_{s}$ being the $s$-wave scattering length of
inter-atomic collisions, is the strength of the contact nonlinearity, which
may be tuned by the Feshbach resonance \cite{Chin2010,Courteille1998}. The
coupling coefficient of the dipole-dipole interaction (DDI) is $\kappa =\mu
_{0}\mu ^{2}/4\pi $, where $\mu _{0}$ and $\mu $ are the vacuum permeability
and atomic magnetic moment of the atom. The coefficient in front of the LHY
term is $\gamma =\left( 32ga_{s}^{3/2}/3\sqrt{\pi }\right) \left(
1+3\epsilon _{\mathrm{dd}}^{2}/2\right) $ \cite%
{Fischer2006,Lima2011,Wachtler2016}, where the relative DDI strength $%
\epsilon _{\mathrm{dd}}\equiv a_{\mathrm{dd}}/a_{s}$ is determined by the
dipole scattering length, $a_{\mathrm{dd}}=\mu _{0}\mu ^{2}m/12\pi \hbar $
\cite{Schmitt2016}. The DDI potential is $U_{\mathrm{dd}}(\mathbf{r}-\mathbf{%
r^{\prime }})=\left( 1-3\cos ^{2}\Theta \right) /|\mathbf{r}-\mathbf{%
r^{\prime }}|^{3}$ \cite{Lahaye2009,Stuhler2005}, where $\cos ^{2}\Theta
=\left( x-x^{\prime }\right) ^{2}/|\mathbf{r}-\mathbf{r^{\prime }}|^{2}$,
and coefficient $\lambda _{3}$ represents the three-body losses.

Disregarding the losses, the stationary solutions with chemical potential $%
\Omega $ are looked for in the usual form, $\psi (\mathbf{r},t)=\phi (%
\mathbf{r})e^{-i\Omega t/\hbar }$, with a stationary wave function $\phi (%
\mathbf{r})$. Equation (\ref{GP-LHY}) with $\Lambda _{3}=0$ conserves the
total atom number, $N=\int |\psi (\mathbf{r})|^{2}d\mathbf{r}$, energy, $%
E=\int d\mathbf{r}\left[ \frac{\hbar ^{2}}{2m}|\nabla \psi |^{2}+\frac{1}{2}%
g|\psi |^{4}+\frac{1}{2}\kappa |\psi |^{2}\int U_{\mathrm{dd}}(\mathbf{r}-%
\mathbf{r^{\prime }})|\psi (\mathbf{r^{\prime }})|^{2}d\mathbf{r^{\prime }}%
\notag\right. \newline
\left. +\frac{2}{5}\gamma |\psi |^{5}\right] $, and momentum (here we
consider quiescent modes, with zero momentum).

3D-AVQD solutions with integer vorticity $S$ can be produced in the
numerical form by means of the imaginary-time-integration method \cite%
{Chifalo2000,Jianke2008,Bao}, initiated with an anisotropic input,
\begin{equation}
\phi ^{(0)}(x,y,z)=A\tilde{r}^{S}\exp \left( -\alpha _{1}\tilde{r}%
^{2}-\alpha _{2}z^{2}+iS\tilde{\theta}\right) ,  \label{ansatz}
\end{equation}%
where $A$ and $\alpha _{1,2}$ are positive real constants which determine
widths of the input, and the deformed polar coordinates in the $\left(
x,y\right) $ plane are $\left\{ \tilde{r},\tilde{\theta}\right\} \equiv
\left\{ \sqrt{x^{2}+\beta ^{2}y^{2}},\arctan (\beta y/x)\right\} $ with an
anisotropy factor $\beta >1$. In this work, we select parameters of the BEC
of dysprosium, $^{164}$Dy, which has a significant dipole scattering length,
$a_{\mathrm{dd}}=131a_{0}$ ($a_{0}$ is the Bohr radius) \cite{Schmitt2016}.
The control parameters of the system are $N$ and $a_{s}$.

\begin{figure}[h]
{\includegraphics[width=0.99\columnwidth]{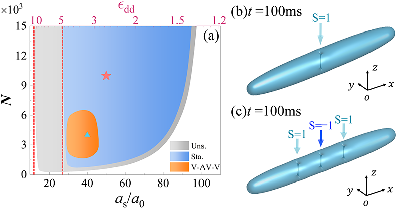}}
\caption{(a) The areas of stable and unstable 3D-AVQDs, as well as stable
vortex-antivortex-vortex bound-states in the plane of $(N,a_{s})$, which
populate the blue, gray and orange areas, respectively. The left dashed
vertical line, at $a_{s}=12a_{0}$, is the existence boundary for the
3D-AVQDs, and the right vertical line, at $a_{s}=27a_{0}$, is their
stability boundary. Red-font numbers, attached to the upper axis, are values
of $\protect\epsilon _{\mathrm{dd}}$. (b-c) Typical examples of a stable
3D-AVQD and vortex-antivortex-vortex bound state with $%
(N,a_{s})=(10^{4},50a_{0})$ and $(4000,40a_{0})$, corresponding to the red
star and blue triangle, respectively, in panel (a). They survive the
perturbed evolution in the course of $100$ ms, at least.}
\label{area}
\end{figure}

The stability of the numerically obtained 3D-AVQDs solutions of Eq. (\ref%
{GP-LHY}) with $S=1$ was tested by real-time simulations of the perturbed
evolution. The numerically found stability area for them in the $(N,a_{s})$
plane is plotted in Fig. \ref{area}(a), with a typical example of a stable
3D-AVQD shown in Fig. \ref{area}(b). The average atomic density of this
state is $140\times 10^{20}$ atoms/m$^{3}$, in agreement with the estimate
in Ref \cite{Ferrier-Barbut2016}. In the simulations, stable 3D-AVQDs, which
populate the blue areas in Fig. \ref{area}(a), maintain their integrity
during a sufficient long time (at least, $\sim 100$ ms), which is longer
than the levitation time ($\sim $ 90 ms) in the experiment \cite{Schmitt2016}%
. On the other hand, the unstable 3D-AVQDs [in the gray area in Fig. \ref%
{area}(a)] spontaneously transform into ground-state QDs after a few
milliseconds. It is thus observed that 3D-AVQDs exist at $a_{s}>12a_{0}$,
and they are stable at $a_{s}>27a_{0}$.

In the 2D geometry, particular \emph{stable} bound-states with a
vortex-antivortex-vortex structure were revealed \cite{Li2023}. Remarkably,
similar bound-states can be created in the current 3D setting too, by means
of input {\small \
\begin{eqnarray}
\phi ^{(0)} &=&\sum_{+,-}A_{\pm }\tilde{r}_{\pm }\exp \left( -\alpha _{1}%
\tilde{r}_{\pm }^{2}-\alpha _{2}z^{2}+i\tilde{\theta}_{\pm }\right)  \notag
\\
&&+A\tilde{r}\exp \left( -\alpha _{1}\tilde{r}^{2}-\alpha _{2}z^{2}-i\tilde{%
\theta}\right) .  \label{3pivotansatz}
\end{eqnarray}%
}Here, $A_{\pm }>0$ and $\alpha _{1,2}>0$ are real constants, $\tilde{r}%
_{\pm }\equiv \sqrt{(x\pm x_{0})^{2}+\beta ^{2}y^{2}}$, $\tilde{\theta}_{\pm
}\equiv \arctan \left[ \beta y/(x\pm x_{0})\right] $, and $x_{0}$ is an
appropriately chosen separation, cf. Eq. (\ref{ansatz}). A typical example
of such a stable composite QD with average density $200\times 10^{20}$
atoms/m$^{3}$ is displayed in Fig. \ref{area}(c). They are stable in the
orange area in the plane of $(N,a_{s})$, which is embedded in the broader
stability region of the regular 3D-AVQDs, see Fig. \ref{area}(a). It is seen
that the stable three-pivot vortex bound-states exist in the region of $%
30<a_{s}/a_{0}<45$ and $1800<N<6400$.

\begin{figure}[h]
{\includegraphics[width=0.99\columnwidth]{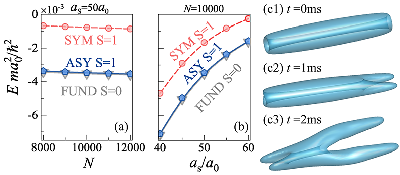}}
\caption{(a-b) The total energy ($E$) of the fundamental QDs (the grey curve
labeled FUND), 3D-AVQDs (the blue curve labeled ASY), and isotropic vortical
QDs (the red curve labeled SYM) versus $N$ (a) and $a_{s}$ (b). (c1-c3) The
unstable evolution of a SYM vortex QD with $(N,a_{s})=(10^{4},50a_{0})$
illustrated by its density profiles at $t=0$ ms, $1$ ms, and $2$ ms,
respectively.}
\label{comparison}
\end{figure}

As mentioned above, the vortex states with the vorticity axis parallel to
the polarization of the dipoles [see \textquotedblleft Type 1" in Fig. \ref{example}] are completely unstable. Because these
solutions are axially symmetric, they are marked SYM in Fig. \ref{comparison}%
. The 3D-AVQD solutions obtained here are anisotropic, therefore they are
marked by the ASY label. Figure \ref{comparison} displays the comparison of
the total energy between the isotropic and anisotropic species of the vortex
solutions. The energy of the fundamental (zero-vorticity) QDs, marked by
FUND, is also included, as a reference. Figures \ref{comparison}(a,b) show
that the unstable SYM vortex QDs have the highest energy (which is a natural
explanation for their instability), while stable ASY vortex states have a
lower energy, which is almost identical to that of the fundamental QDs.

For the SYM type of the vortex QDs, the void around the long axis implies
the removal of a long tube filled by dipoles chiefly featuring attractive
DDIs, i.e., the removal of the negative interaction energy, which causes
them to have higher actual energy values, in accordance with Fig. \ref%
{comparison}(a,b). A typical example of the evolution of the SYM vortex-QD
is displayed in Fig. \ref{comparison}(c1-c3), which demonstrates the
instability-induced splitting. These results agree with the instability of
the isotropic vortex solitons that was reported in Ref. \cite{Cidrim2018}.
On the other hand, the stability of the ASY type is feasible because the
corresponding axial void removes a tube filled by dipoles chiefly featuring
repulsive DDIs with the positive energy, thus producing lower actual energy
values, as corroborated by Fig. \ref{comparison}(a,b). Additional analysis
has demonstrated that the application of the imaginary-time-integration
method to Eq. (\ref{GP-LHY}) does not generate 3D-AVQD solutions with
multiple vorticity, $S\geq 2$.

\begin{figure}[h]
{\includegraphics[width=0.99\columnwidth]{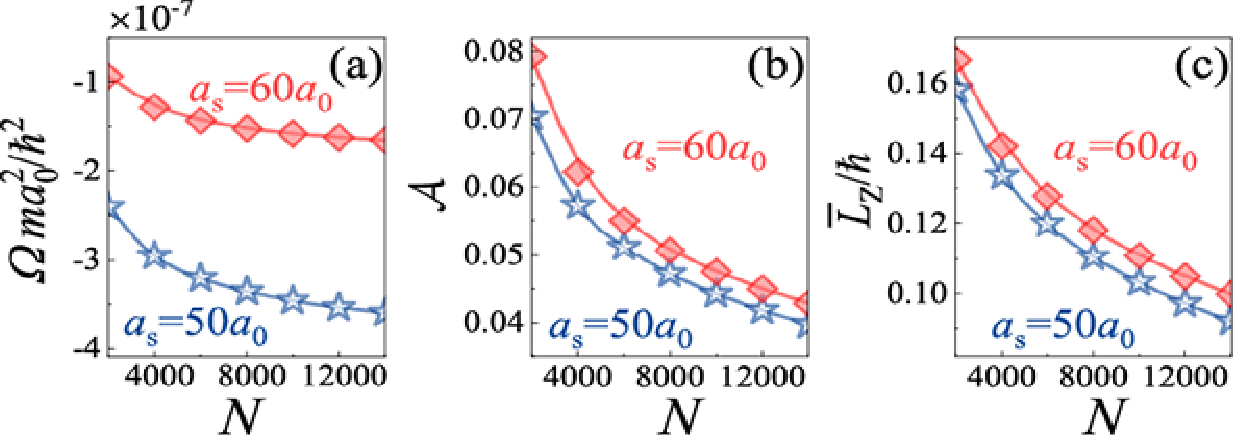}}
\caption{(a) The chemical potential ($\Omega $), (b) ellipticity ($\mathcal{A%
}$), and (c) angular momentum ($\bar{L}_{z}$) (see Eq. (\protect\ref%
{definition})) versus $N$, for $a_{s}=50a_{0}$ and $60a_{0}$ (the chains of
blue stars and red rhombuses, respectively).}
\label{char}
\end{figure}

To present systematic results for the 3D-AVQDs, we define their ellipticity $%
\mathcal{A}$ and normalized angular momentum $\bar{L}_{z}$:
\begin{equation}
\mathcal{A}=\frac{D_{\mathrm{S}}}{D_{\mathrm{L}}},\quad \bar{L}_{z}=\int
\frac{\phi ^{\ast }\hat{L}_{z}\phi }{N}d\mathbf{r},  \label{definition}
\end{equation}%
where $D_{\mathrm{S}}$ and $D_{\mathrm{L}}$ are, respectively, the short and
long axes of the QDs, and $\hat{L}_{z}=i\hbar (y\partial _{x}-x\partial
_{y}) $ is the operator of the $z$ components of the angular momentum.
Dependences of the chemical potential, ellipticity, and angular momentum on
the number of atoms, for two different values of $a_{s}$, are produced in
Fig. \ref{char}.

In Fig. \ref{char}(a), the chemical potential $\Omega $ satisfies the
Vakhitov-Kolokolov criterion, $d\Omega /dN<0$, which is the well-known
necessary stability condition for self-trapped modes \cite{VK1973,Berge1998}%
. A basic feature of QDs is their incompressibility. This implies that the
average density of the droplets cannot exceed a maximum value \cite%
{Petrov2015}, which leads to flat-top QDs' shape. Thus, the volume of the
QDs increases linearly with the growth of the number of atoms. Further, due
to the strong DDI anisotropy, the increase of the volume is mostly
represented by the extension along the $x$-axis, leading to the decrease of
the ellipticity [see Eq. (\ref{definition})] in Fig. \ref{char}(b). As the
internal vorticity is mainly concentrated at the center of the droplet, Fig. %
\ref{char}(c) shows that larger values of norm correspond to longer droplets
and lower values of the angular momentum. Moreover, figures \ref{char}(b,c)
reveal that $\bar{L}_{z}/\hbar =2\mathcal{A}$, which coincides with the
relation found for strongly anisotropic 2D-AVQDs \cite{Li2023}.
%On the other hand, even when $a_{s}$ increases, the most obvious effect is to increase the droplets' volume, but in reality, the changes of these parameters with $a_{s}$ are not entirely consistent with the changes with total atom number because $a_{s}$ affects the interaction strength of DDIs and quantum fluctuation effect at the same time.

\begin{figure}[h]
{\includegraphics[width=0.9\columnwidth]{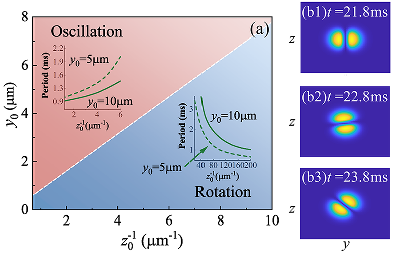}}
\caption{(a) The oscillation and rotation regions in the plane of the
torque's parameters, for $(N,a_{s})=(10^{4},50a_{0})$. The insets show the
oscillation and rotation periods vs. $z_{0}^{-1}$ for fixed $y_{0}=5$ and $%
10 $ $\mathrm{\protect\mu }$m (the dashed and solid lines, respectively).
(b1-b3) Plots of the cross-section density in the $(y,z)$ plane illustrating
the robust rotation of the 3D-AVQD around the $x$-axis, with period $5.5$
ms, initiated by the torque with $(z_{0},y_{0})=(0.05,5)$ $\mathrm{\protect%
\mu }$m.}
\label{torque}
\end{figure}

The shape of the 3D-AVQDs suggests a possibility to set it in rotation
around an axis perpendicular to the vorticity vector. To this end, a torque
was applied around the $x$-axis, multiplying the established 3D-AVQD by the
phase factor $\exp [i(z/z_{0})\tanh (y/y_{0})]$, i.e., adding an $x$%
-component of the angular momentum to the original $z$-component, cf. Ref.
\cite{Kartashov2014}. Here, $z_{0}$ and $y_{0}$ are length scales, which
define the strength of the torque. Simulations reveal oscillations or
rotation of the 3D-AVQDs around the $x$-axis, depending on values of $z_{0}$
and $y_{0}$. The weak torque, corresponding to large $(z_{0}$ and $y_{0})$,
induces oscillations, whose period increases with the decrease of $z_{0}$
and $y_{0}$. Divergence of the oscillation period implies a transition to
the rotation, caused by a sufficiently strong torque (see Movies I and II
in Supplemental Material). The rotation speed increases with the further
decrease of $y_{0}$ and $z_{0}$, as the torque is made still stronger.
Figure \ref{torque}(a) displays the oscillation and rotation regions in the
plane of $\left( z_{0}^{-1},y_{0}\right) $ for $(N,a_{s})=(10^{4},50a_{0})$.
The border between these dynamical regimes is fitted by $%
y_{0}=Z_{0}^{2}/z_{0}+Y_{0}$, with $Z_{0}\approx 0.88$ $\mathrm{\mu }$m and $%
Y_{0}\approx 0.06$ $\mathrm{\mu }$m. This relation is explained by the fact
that, for $|y|\lesssim y_{0}$, the torque's phase, $\approx yz/\left(
y_{0}z_{0}\right) $, is determined solely by product $y_{0}z_{0}$. Periods
of the oscillations and rotation are displayed, as functions of $z_{0}^{-1}$%
, by insets in the respective regions. A typical example of the stable
rotation is presented in Fig. \ref{torque}(b1-b3). The rotation picture is
the same as produced by the stationary solution of Eq. (\ref{GP-LHY}) in the
rotating reference frame, which includes term $\omega \hat{L}_{x}\psi $,
where $\hat{L}_{x}=i\hbar (z\partial _{y}-y\partial _{z})$ and $\omega $ is
the rotation frequency.

We have also explored results of the application of the torque around the $y$%
- and $z$-axes, in terms of Fig. \ref{example}. In the former case, the
torque drives a complex dynamical regime: the prolate QD features
oscillations in the $\left( z,x\right) $ plane, simultaneously with
irregular rotation around the $x$ axis (not around the $y$ direction), as
shown by Movie III in Supplemental Material. Lastly, the application of a
weak torque along the $z$ direction initiates oscillations of the prolate
vortex soliton in the $\left( x,y\right) $ plane (see Movie IV in
Supplemental Material), while a stronger torque leads to its splitting, the
boundary between the two regimes being $x_{0}=Y_{0}^{2}/y_{0}+X_{0}$, where $%
Y_{0}\approx 0.67$ $\mathrm{\mu }$m and $X_{0}\approx -0.5$ $\mathrm{\mu }$%
m, in terms of the torque's spatial scales.

\begin{figure}[h]
{\includegraphics[width=0.9\columnwidth]{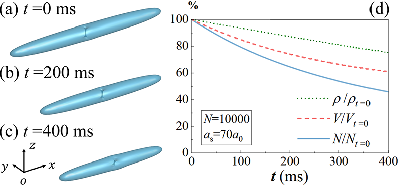}}
\caption{(a-c) The real-time evolution of the droplet with three-body losses
for $N=10^{4}$ and $a_{s}=70a_{0}$. (d) The residual ratio of $N$, $V$, and $%
\protect\rho $ vs. time.}
\label{TL}
\end{figure}

Finally, it is imperative to consider the effect of three-body
losses, characterized by coefficient ${\Lambda }${$_{3}=1.25\times 10^{-41}$
m$^{6}$s$^{-1}$ in Eq. (\ref{GP-LHY}) \cite{Schmitt2016}. In general, losses
may attenuate instabilities for fundamental (zero-vorticity) states, but
this is not applicable to vortex QDs, whose stability is determined by the
equilibrium value of the density. We observe that the scattering length $%
a_{s}$ significantly affects the loss effect. Notably, for $N=10^{4}$, if $%
a_{s}$ is smaller than a critical value, $66a_{0}$, the losses drive rapid
degeneration of the initial vortex QD into a fundamental (zero-vorticity)
state in the course of $<100$ ms (see Movie V in Supplemental Material). The
residual state survives much longer, which implies that the QD's topological
structure is especially vulnerable to the loss effect. However, at $%
a_{s}>66a_{0}$, the robustness is much improved. For example, as shown in
Fig. \ref{TL}(a-c), the vortex-QD with $a_{s}=70a_{0}$ retains its
topological charge for more than 400 ms (see Movie VI in Supplemental
Material). As shown in Fig. \ref{TL}(d), in the latter case the total atom
number $N$, effective volume $V=(\int |\psi |^{2}d\mathbf{r})^{2}/\int |\psi
|^{4}d\mathbf{r}$, and density $\rho =N/V$ of the QD decrease slowly,
demonstrating that the losses are not a fatal factor.}

%There may be at least two approaches to create the 3D-AVQDs in experiment. One is suggested in the 2D case \cite{Li2023}. This approach utilizes the unstable dipole modes, which spontaneously transform into vortex modes rapidly during real-time evolution. This process can be also found in the simulations of present setting. Another more common option is to first prepare the BEC of $^{164}$Dy and shape it into an elongated shape along the magnetic field direction using an additional optical trap and directly imprint the desired phase pattern by using vortical laser beams \cite{Shen2019}. Then turn off the traps after the creation of vortices.

\emph{Conclusion} We have predicted the existence of stable AVQDs
(anisotropic vortex quantum droplets) in 3D dipolar BECs, with the strong
anistropy imposed by the orthogonality of the vorticity and polarization of
atomic magnetic moments. While isotropic vortex solitons in dipolar BEC are
completely unstable, we have identified a vast stability region of 3D-AVQDs
in the system's parameter space. The existence of stable composite states
with the vortex-antivortex-vortex structure is demonstrated as well, and
their stability area is identified. Essential characteristics of the 3D
AVQDs, including the chemical potential, aspect ratio, and angular momentum,
are presented as functions of control parameters. Furthermore, we have
demonstrated that the application of the torque perpendicular to the
vorticity axis initiates robust intrinsic oscillations or rotation of the
3D-AVQDs. The dependence of the oscillation and rotation periods on
parameters of the torque have been found. The persistence of the 3D-AVQDs
under the action of three-body losses was analyzed too, demonstrating that
the topological structure (vorticity) is retained by the 3D AVQD for a
sufficiently long time when the scattering length exceeds a critical value.

As an extension of the present analysis, it may be relevant to look for more
complex bound states of AVQDs, and to study a two-component version of the
model, cf. Refs. \cite{Gammatwocomponent,Bisset2021,Smith2021}. Another
relevant problem is to add an ingredient (probably, an external potential)
that may help to stabilize higher-order anisotropic vortex solitons, with $%
S\geq 2$.

%\begin{acknowledgments}

\begin{acknowledgments}
Authors appreciate valuable discussions with Profs. Zhenya Yan, G. E.
Astrakharchik, Yongchang Zhang, and Dr. Xizhou Qin. This work was supported by NNSFC (China)
through Grants No. 12274077, 12305013, 11874112, 11905032, the Natural Science Foundation of Guangdong province through
Grant No. 2024A1515030131, 2023A1515010770, the Research Fund of Guangdong-Hong
Kong-Macao Joint Laboratory for Intelligent Micro-Nano Optoelectronic
Technology through grant No.2020B1212030010. The work of B.A.M. is
supported, in part, by the Israel Science Foundation through grant No.
1695/22.
\end{acknowledgments}

\end{document}